# Spectral sum rules and search for periodicities in DNA sequences

V.R. Chechetkin[*]

*Theoretical Department of Division for Perspective Investigations, Troitsk Institute of Innovation and Thermonuclear Investigations (TRINITI), Troitsk,*
*142190 Moscow Region, Russia*

**Abstract**

Periodic patterns play the important regulatory and structural roles in genomic DNA sequences. Commonly, the underlying periodicities should be understood in a broad statistical sense, since the corresponding periodic patterns have been strongly distorted by the random point mutations and insertions/deletions during molecular evolution. The latent periodicities in DNA sequences can be efficiently displayed by Fourier transform. The criteria of significance for observed periodicities are obtained via the comparison versus the counterpart characteristics of the reference random sequences. We show that the restrictions imposed on the significance criteria by the rigorous spectral sum rules can be rationally described with De Finetti distribution. This distribution provides the convenient intermediate asymptotic form between Rayleigh distribution and exact combinatoric theory.

## 1. Introduction

Tandem repeats and scattered DNA repeats play important roles in the structural organization of chromosomes and regulatory mechanisms in a variety of organisms [1]. Commonly, however, the periodic patterns in DNA sequences are strongly randomized by the point mutations and insertions/deletions during molecular evolution. In this case discrete Fourier transform provides the efficient tool to study the intricate relationships between structure and function from the underlying quasiperiodic patterns in DNA or protein sequences [2–14] (for a review see, e.g., [15, 16]). The correct inference about latent periodicities revealed by Fourier analysis needs the applications of the proper significance criteria. The random sequences of the same molecular composition serve as a reference standard for the assessment of observed regularities. If the length $M$ of a sequence is long enough (or the relative variations of statistical characteristics $\sim M^{-1/2}$ are small), the corresponding criteria satisfy the property of self-averaging and may be applied to a given sequence under investigation. In this work we will discuss how the rigorous spectral sum rules affect the significance criteria for the periodicities in DNA sequences. Among the applications of the theory, we should mention the search for long periods via the sums of equidistant harmonics, the analysis of structural subharmonics, the study of correlations between different periodicities, and the assessment of variations in the spectral entropy. The approach can be generalized to the arbitrary symbolic sequences and discrete Fourier transform.

---

[*] *E-mail addresses:* chechet@biochip.ru and vladimir_chechet@mail.ru.

**2. Fourier transform of DNA sequences. Rayleigh and De Finetti distributions**

DNA sequences are composed of the nucleotides of four types A, C, G, or T. The Fourier harmonics corresponding to the nucleotides of type α (where α is A, C, G, or T) in a sequence of length $M$ are defined as

$$\rho_\alpha(q_n) = M^{-1/2} \sum_{m=1}^{M} \rho_{m,\alpha} e^{-iq_n m}, \quad q_n = 2\pi n/M, \quad n = 0, 1, ..., M-1 \qquad (1)$$

where $\rho_{m,\alpha}$ indicates the positions occupied by the nucleotides of type α, $\rho_{m,\alpha} = 1$ if the nucleotide of type α occupies the $m$-th site and 0 otherwise. The amplitudes of Fourier harmonics (or structure factors) are expressed as

$$F_{\alpha\alpha}(q_n) = \rho_\alpha(q_n) \rho_\alpha^*(q_n) \qquad (2)$$

Here asterisk denotes the complex conjugation. Due to symmetry property

$$F_{\alpha\alpha}(q_n) = F_{\alpha\alpha}(2\pi - q_n) \qquad (3)$$

the spectra for structure factors can be restricted to their left halves from $n = 1$ to

$$N = [M/2] \qquad (4)$$

where the brackets denote the integer part of the quotient. The structure factors obey the rigorous sum rule [2, 3, 15]

$$\sum_{n=0}^{M-1} F_{\alpha\alpha}(q_n) = \sum_{m=1}^{M} \rho_{m,\alpha} \rho_{m,\alpha} = N_\alpha \qquad (5)$$

where $N_\alpha$ is the total number of nucleotides of type $\alpha$ in a sequence of length $M$. The zeroth harmonics depending only on the nucleotide composition do not contain structural information and should be excluded. Introducing the normalization

$$f_{\alpha\alpha}(q_n) = F_{\alpha\alpha}(q_n) / \overline{F}_{\alpha\alpha}; \quad \overline{F}_{\alpha\alpha} = N_\alpha (M - N_\alpha) / M(M-1) \qquad (6)$$

the sum rule for harmonics with $n \neq 0$ can be presented in the form

$$\sum_{n=1}^{M-1} f_{\alpha\alpha}(q_n) = M - 1 \qquad (7)$$

The calculation of various moments for the structure factors in random sequences is reduced to the combinatoric averaging of products $\prod_{\alpha \in A,C,G,T} \prod_{k_\alpha=1}^{n_\alpha} \rho_{m_{k_\alpha},\alpha}$ at the given nucleotide composition [3, 15]. As the correlations between structure factors with the different wavenumbers $q_n$ are small, $\sim 1/N$, the multivariable probability for half-spectrum is given in the main approximation by the product of Rayleigh distributions for particular harmonics

$$P(f_{\alpha\alpha}(q_1) \geq f_1, ..., f_{\alpha\alpha}(q_N) \geq f_N) = \exp(-f_1 - ... - f_N) \qquad (8)$$

For distribution (8) the sum rule (7) is fulfilled only in average. This appears to be insufficient for a part of applications.

Below we will use the sum rule for independent structure factors in the form

$$\sum_{n=1}^{N} f_{\alpha\alpha}(q_n) = N \qquad (9)$$

which is exact for odd *M* and approximate for even *M*. Let us reciprocate the situation and search for the corresponding distribution starting directly from the sum rule (9) and implying uniformity for different structure factors $f_{\alpha\alpha}(q_n)$. This means that multivariable probability density should be proportional to δ-function

$$p_N(f_{\alpha\alpha}(q_1); ...; f_{\alpha\alpha}(q_N)) \propto \delta(1 - f_{\alpha\alpha}(q_1)_1/N - ... - f_{\alpha\alpha}(q_N)/N) \qquad (10)$$

Here and below the capital letter *P* always denotes probability, whereas the small letter *p* denotes the corresponding probability density. Integrating (10) over region $N \geq f_{\alpha\alpha}(q_n) \geq f_n; n = 1,...,N$ and choosing proper normalization factor provides De Finetti distribution [17]

$$P(f_{\alpha\alpha}(q_1) \geq f_1; ...; f_{\alpha\alpha}(q_N) \geq f_N) = \left(1 - f_1/N - ... - f_N/N\right)_+^{N-1} \qquad (11)$$

where $x_+ = x$, if $x > 0$ and $x_+ = 0$, if $x \leq 0$. Alternatively, De Finetti distribution can also be derived as conditional probability for the structure factors with identical Rayleigh distributions subject to the constraint (9). In the limit of large *N* De Finetti distribution (11) tends to Rayleigh distribution (8) playing the universal role in the spectral analysis.

**3. One-harmonic distribution**

One-harmonic distribution and probability density are obtained from the general distribution (11) as

$$P(f_{\alpha\alpha}(q_1) \geq f; f_{\alpha\alpha}(q_2) \geq 0; ...; f_{\alpha\alpha}(q_N) \geq 0) = \left(1 - f/N\right)_+^{N-1};$$
$$p(f) = \left(1 - f/N\right)_+^{N-2}(N-1)/N \qquad (12)$$

This gives the moments

$$<f^k> = N^k \frac{k!(N-1)!}{(k+N-1)!} \qquad (13)$$

(henceforth the angular brackets denote the averaging over probability density), in particular,

$$<f> = 1; <f^2> = \frac{2N}{N+1}; \sigma^2(f) = \frac{N-1}{N+1} \tag{14}$$

The corresponding distributions and moments for extreme values (minimum or maximum of structure factors in the whole half-spectrum) are expressed as (cf. [17])

$$P(\min f_{\alpha\alpha}(q_n) > f) = (1-f)^{N-1};$$
$$<\min f_{\alpha\alpha}(q_n)> = \frac{1}{N}; \quad \sigma^2(\min f_{\alpha\alpha}(q_n)) = \frac{N-1}{N^2(N+1)} \tag{15}$$

$$P(\max f_{\alpha\alpha}(q_n) \leq f) = \sum_{k=0}^{N}(-1)^k \binom{N}{k}(1-kf/N)_+^{N-1};$$
$$<\max f_{\alpha\alpha}(q_n)> = \sum_{k=1}^{N}\frac{1}{k}; \quad \sigma^2(\max f_{\alpha\alpha}(q_n)) = \frac{N}{(N+1)}\left[\left(\sum_{k=1}^{N}\frac{1}{k^2}\right) - \frac{1}{N}\left(\sum_{k=1}^{N}\frac{1}{k}\right)^2\right] \tag{16}$$

The pronounced peaks in the whole genomic spectra should be compared with the singular outbursts in random spectra given by Eq. (16) rather than with the threshold for one-harmonic probability (12). The relevant criteria for the maximum from $W < N$ harmonics may be obtained quite similarly,

$$P(\max\{f_i\}_{i=1}^{W} \leq f) = \sum_{k=0}^{W}(-1)^k \binom{W}{k}(1-kf/N)_+^{N-1};$$
$$<\max\{f_i\}_{i=1}^{W}> = \sum_{k=1}^{W}\frac{1}{k}; \quad \sigma^2(\max\{f_i\}_{i=1}^{W}) = \frac{N}{(N+1)}\left[\left(\sum_{k=1}^{W}\frac{1}{k^2}\right) - \frac{1}{N}\left(\sum_{k=1}^{W}\frac{1}{k}\right)^2\right] \tag{17}$$

## 4. Two-harmonic distribution and spectral entropy

The two-harmonic distribution and probability density

$$P(f_{\alpha\alpha}(q_1) \geq f_1; f_{\alpha\alpha}(q_2) \geq f_2; ...; f_{\alpha\alpha}(q_N) \geq 0) = (1 - f_1/N - f_2/N)_+^{N-1};$$
$$p(f_1; f_2) = (1 - f_1/N - f_2/N)_+^{N-3}(N-1)(N-2)/N^2 \tag{18}$$

can be applied to the assessment of pair correlations. The simplest correlator is given by

$$<f_{\alpha\alpha}(q_1)f_{\alpha\alpha}(q_2)> - <f_{\alpha\alpha}(q_1)><f_{\alpha\alpha}(q_2)> = -1/(N+1) \tag{19}$$

Another application of distribution (18) refers to the spectral entropy [3, 15]

$$S_\alpha = -\sum_{n=1}^{M-1} f_{\alpha\alpha}(q_n) \ln f_{\alpha\alpha}(q_n) \qquad (20)$$

The spectral entropy characterizes the general uniformity of a spectrum or the general level of regularity of DNA sequence (for particular applications of spectral entropy to genetic problems see [18–20]). In the limit of long sequences the calculation yields

$$<S_\alpha> = -0.4227...(M-1); \quad <(\Delta S_\alpha)^2> = 0.5797...(M-1) \qquad (21)$$

The normalized deviations $(<S_\alpha> - S_\alpha)/<(\Delta S_\alpha)^2>^{1/2}$ obey Gaussian statistics and serve for the quantitative assessment of general regularity of DNA sequence under study. The higher normalized deviations correspond to the higher structural regularity of a sequence.

## 5. Distributions of sums

Generally, the latent periodicities in DNA sequences should be searched through the sums of equidistant peaks in Fourier spectrum [4, 13, 15]. The relevant characteristics for the sum of $k$ structure factors in random sequences can be assessed with distributions

$$P(S_k > S) = \sum_{j=0}^{k-1} \binom{N-1}{j} \left(\frac{S}{N}\right)^j \left(1 - \frac{S}{N}\right)_+^{N-j-1};$$

$$p_k(S) = \frac{(N-1)}{N} \binom{N-2}{k-1} \frac{S^{k-1}}{N^{k-1}} \left(1 - \frac{S}{N}\right)_+^{N-k-1} \qquad (22)$$

This yields the moments

$$<S_k^m> = N^m \frac{(N-1)!}{(k-1)!} \frac{(k+m-1)!}{(N+m-1)!} \qquad (23)$$

The corresponding mean sum and its variations are equal to

$$<S_k> = k; \quad \sigma^2(S_k) = k\frac{N-k}{N+1} \tag{24}$$

At $k = N$ the variations are absent $<S_N^m> = N^m$ in accordance with rigorous sum rule (9). As underlying periodicities in DNA sequences are superimposed, the equidistant peaks in Fourier spectrum may slightly variate. The respective criteria can easily be derived by combining (17) and (24) (cf. also [4, 15]).

**6. Correlation coefficients**

Pearson correlation coefficient between structure factors with different wavenumbers in random sequences can be assessed as

$$<k(f_{\alpha\alpha}(q_n)|f_{\alpha\alpha}(q_{n'}))> \equiv <k_{\alpha\alpha}(n|n')> \approx$$
$$\frac{<f_{\alpha\alpha}(q_n)f_{\alpha\alpha}(q_{n'})> - <f_{\alpha\alpha}(q_n)><f_{\alpha\alpha}(q_{n'})>}{<\sigma(f_{\alpha\alpha}(q_n))><\sigma(f_{\alpha\alpha}(q_{n'}))>} = -1/(N-1) \approx -1/N \tag{25}$$

The corresponding correlation coefficient for the nucleotides of different types may be derived from the sum rule

$$\sum_{n=1}^{N} f_{\alpha\alpha}(q_n) + \sum_{n=1}^{N} f_{\beta\beta}(q_n) = 2N \tag{26}$$

Squaring and averaging (26) yields equality

$$\sum_{n}\left(<f_{\alpha\alpha}(q_n)f_{\beta\beta}(q_n)> - <f_{\alpha\alpha}(q_n)><f_{\beta\beta}(q_n)>\right) +$$
$$\sum_{n \neq n'}\left(<f_{\alpha\alpha}(q_n)f_{\beta\beta}(q_{n'})> - <f_{\alpha\alpha}(q_n)><f_{\beta\beta}(q_{n'})>\right) = 0 \tag{27}$$

Then, just in lines with (25) one obtains

$$<k(f_{\alpha\alpha}(q_n)|f_{\beta\beta}(q_{n'}))> \equiv <k_{\alpha\beta}(n|n')> \approx -<k_{\alpha\beta}(n|n)>/N \tag{28}$$

The correlation coefficient for structure factors with coincident wavenumbers was derived earlier [3, 15]

$$<k_{\alpha\beta}(n|n)> = N_\alpha N_\beta /(M-N_\alpha)(M-N_\beta) \qquad (29)$$

If the correlation coefficients are assessed via the averaging over a set of $P$ random sequences, the corresponding standard deviations from the mean values would be about $1/\sqrt{P}$ in the limit of large $P$. Correlations (25) and (28) can also be derived within the frameworks of combinatoric theory, but in the latter case the calculations are incomparably more tedious.

## 7. Conditional probability

Let a spectrum display some singular peaks. This raises the problem about the significance of remaining harmonics after subtraction of singular values from the sum rule (9). Let the sum of remaining $N'$ harmonics be equal to $S'$. Then, it is easy to see that all the criteria will hold after rescaling $f \to (S'/N')f$, which solves the posed conditional problem.

## 8. Generalizations

Primarily, we were interested in the study of periodicities in DNA sequences. Nevertheless, the derived criteria may be efficiently applied to the temporal physical data as well. The discrete Fourier transform for the real temporal data, $u(t_m) \equiv u_m$, is defined as

$$u(\omega_n) = M^{-1/2} \sum_{m=1}^{M} u_m e^{-i\omega_n m \Delta t}, \quad \omega_n = 2\pi n/M\Delta t, \quad n = 0,1,...,M-1, \qquad (30)$$

$$F(\omega_n) = u(\omega_n)u^*(\omega_n) \qquad (31)$$

where $\Delta t$ is sampling time. The mean spectral intensity is connected with dispersion $\sigma(u)$ as

$$(M-1)^{-1} \sum_{n=1}^{M-1} F(\omega_n) = \sigma^2(u) \qquad (32)$$

where

$$\sigma^2(u) = (M-1)^{-1} \sum_{m=1}^{M} (u_m - \bar{u})^2; \quad \bar{u} = M^{-1} \sum_{m=1}^{M} u_m \tag{33}$$

Defining normalized intensities $f(\omega_n) = F(\omega_n)/\sigma^2(u)$ again reduces sum rule (32) to (9). The significance of underlying periodicities can be assessed by comparison with the statistical characteristics of random data having the same dispersion $\sigma(u)$ as the data under study.

**Acknowledgements**